\documentclass[showpacs,12pt,twocolumn]{iopart}
\usepackage{graphicx}
\usepackage{graphics}
\usepackage{epsfig}
\usepackage{CJK}
\usepackage{color}

\begin{document}

\title[]{Large magnetoresistance in LaBi: origin of field-induced resistivity upturn and plateau in compensated semimetals}

\author{Shanshan Sun$^{1,\dag}$, Qi Wang$^{1,\dag}$, Peng-Jie Guo$^{1,\dag}$, Kai Liu$^{1}$, and Hechang Lei$^{1,*}$}

\address{$^{1}$Department of Physics and Beijing Key Laboratory of Opto-electronic Functional Materials $\&$ Micro-nano Devices, Renmin University of China, Beijing 100872, China}
\ead{hlei@ruc.edu.cn}

\begin{abstract}
The discovery of non-magnetic extreme magnetoresistance (XMR) materials has induced great interests because the XMR phenomenon challenges our understanding of how a magnetic field can alter electron transport in semimetals. Among XMR materials, the LaSb shows XMR and field-induced exotic behaviors but it seems to lack the essentials for these properties. Here, we study the magnetotransport properties and electronic structure of LaBi, isostructural to LaSb. LaBi exhibits large MR as in LaSb, which can be ascribed to the nearly compensated electron and hole with rather high mobilities. More importantly, our analysis suggests that the XMR as well as field-induced resistivity upturn and plateau observed in LaSb and LaBi can be well explained by the two-band model with the compensation situation. We present the critical conditions leading to these field-induced properties. It will contribute to understanding the XMR phenomenon and explore novel XMR materials.

\end{abstract}

%Uncomment for PACS numbers title message
%\pacs{00.00, 20.00, 42.10}
% Keywords required only for MST, PB, PMB, PM, JOA, JOB?
%\vspace{2pc}
%\noindent{\it Keywords}: Article preparation, IOP journals
% Uncomment for Submitted to journal title message
%\submitto{\JPA}
% Comment out if separate title page not required
\maketitle

\section{Introduction}

Magnetoresistance (MR) is the change of electronic resistance with magnetic field. It is not only one of most imporatant topics in condensed matter physics but also vital to practical applications, such as magnetic field sensors, read heads of hard drives, random access memories \cite{Daughton}. In general, the large MR is emergent in magnetic materials, such as giant MR (GMR, MR $\sim$ 10$^{2}$ \%) in metallic thin films and colossal MR (CMR, MR $\sim$ 10$^{4}$ \%) in Cr-based chalcogenide spinels as well as Mn-based perovskites \cite{Baibich,Ramirez,Jin}. In the past several years, a series of non-magnetic metals with extreme MR (XMR) have been discovered, such as $TPn$ ($T$ = Nb and Ta, $Pn$ = P and As) \cite{Huang,Shekhar,LuoY}, Cd$_{3}$As$_{2}$ \cite{LiangT},  NbSb$_{2}$ \cite{Wang}and WTe$_{2}$ \cite{Mazhar}. These discoveries greatly challenge the traditional understanding that the non-magnetic metals usually have a small MR of only about a few percent. Most of novel XMR materials exhibit some common features, such as extremely high mobilities $\mu$, usually $\mu>$ 1$\times$10$^{4}$ cm$^{2}$ V$^{-1}$ s$^{-1}$, and comparable carrier concentrations of electron and hole \cite{Huang,Shekhar,LuoY,LiangT,Mazhar}. Based on these features, several mechanisms have been proposed to explain the origin of XMR in these materials, including topological protection from backscattering mechanism in Dirac and Weyl semimetals (SMs) \cite{LuoY,LiangT}, and electron-hole compensation mechanism \cite{Mazhar}.

Very recently, the XMR materials are extended to the rare-earth-based material LaSb with simple rock-salt structure \cite{Tafti}. It shows XMR at 2 K and 9 T (0.9$\times$10$^{6}$ \%). Moreover, metallic LaSb exhibits the field-induced upturn behavior in resistivity curve appearing at low temperature, followed by a resistivity plateau. Similar phenomena are also observed in several systems, such as NbP and WTe$_{2}$ \cite{Shekhar,Mazhar}. Although the XMR in LaSb and LaBi have been reported in early works \cite{Kasuya1,Kasuya2}, the origin of field-induced upturn behavior and resistivity plateau is not clear. Previous study \cite{Tafti} suggests that the prerequisites for XMR and related field-induced properties seem absent in LaSb, such as the inversion symmetry broken as in Weyl SMs, linear dispersion of band structure near Fermi energy level $E_{F}$ as in Dirac SMs, and perfect electron-hole compensation satisfying the resonance situation as in WTe$_{2}$. Tafti et al proposed that the conducting surface states of a topological insulator protected by time reversal symmetry could be the reason causing these field-induced behaviors in LaSb \cite{Tafti}. But recent angle-resolved photoemission spectroscopy (ARPES) measurement suggests that LaSb is topologically trivial, no Dirac-like surface states are observed \cite{ZengLK}. Moreover, the ARPES results also indicate that the carriers in LaSb are almost compensated. Thus, it is interesting to study whether electron-hole compensation mechanism is valid to explain these exotic phenomena in LaSb.

In order to explore more XMR materials in this family and understand field-induced properties in LaSb as well as other XMR materials, here, we report the detailed study on magnetotransport properties and electronic structure of LaBi single crystals, isostructural to LaSb. We find LaBi exhibits the XMR behavior, consistent with previous results \cite{Kasuya2}. Furthermore, the field-induced resistivity upturn and plateau shown in LaSb are also observed in LaBi. Importantly, the XMR and field-induced phenomena in La$Pn$ can be well understood in the framework of the classical multi-band model. Based on this simple model, we present the key conditions leading to these field-induced behaviors.

\section{Experimental}

Single crystals of LaBi were grown by the In flux. The elements were put into alumina crucible and sealed in quartz ampoule under partial argon atmosphere. The sealed quartz ampoule was heated to and soaked at 1323 K, then slowly cooled down to 1123 K. Finally, the ampoule was decanted with a centrifuge. X-ray diffraction (XRD) of a single crystal were performed using a Bruker D8 X-ray machine with Cu $K_{\alpha}$ radiation. Electrical transport measurements were carried out by using Quantum Design PPMS-14T. The longitudinal and Hall electrical resistivity were performed using a four-probe method on crystals with size of 1$\times$0.7$\times$0.2 mm$^{3}$. The current flows in the $ab$ plane and the magnetic field is perpendicular to current direction. The Hall resistivity was obtained by the difference of transverse resistance measured at the positive and negative fields in order to effectively get rid of the longitudinal resistivity contribution due to voltage probe misalignment, i.e., $\rho_{xy}(\mu_{0}H)=[\rho(+\mu_{0}H)-\rho(-\mu_{0}H)]/2$. The first-principles electronic structure calculations were implemented with the projector augmented wave method \cite{paw} as implemented in the VASP package \cite{vasp}. For the exchange-correlation potential, the generalized gradient approximation of Perdew-Burke-Ernzerhof type \cite{pbe} was used. The kinetic energy cutoff of the plane wave basis was set to be 300 eV. For the Brillouin zone sampling, a 10$\times$10$\times$10 k-point mesh was employed. The spin-orbital-coupling effect was included in the calculations of electronic properties. The maximally localized Wannier functions \cite{Marzari,Souza} was used to calculate the Fermi surfaces (FSs).

\section{Results and discussion}

\begin{figure}[tbp]
\centerline{\includegraphics[scale=0.6]{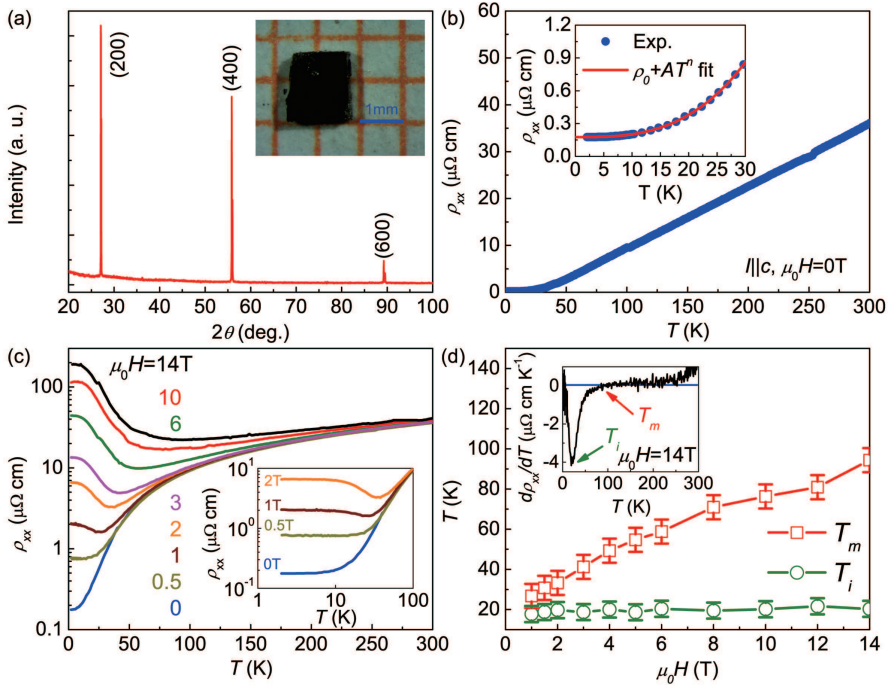}} \vspace*{-0.3cm}
\caption{(a) XRD of a LaBi single crystal at 300 K. Inset: photo of typical LaBi single crystal. (b) Temperature dependence of $\rho_{xx}(T,0)$ for $I\parallel c$. Insets: enlarged parts of $\rho_{xx}(T,0)$ below 20 K. The solid line is the fit using the formula $\rho_{xx}(T,0)=\rho_{0}+AT^{n}$. (c) Temperature dependence of $\rho_{xx}(T, \mu_{0}H)$ at various fields. Inset: enlarged parts of $\rho_{xx}(T, \mu_{0}H)$ at low-temperature region under low fields. (d) Field dependence of $T_{m}$ and $T_{i}$, corresponding to the sign change and the minimum in the $d\rho_{xx}(T, \mu_{0}H)/dT$ curves, respectively. Inset: $d\rho_{xx}(T, \mu_{0}H)/dT$ vs. $T$ at 14 T. The positions of $T_{m}$ and $T_{i}$ are marked by arrows.}
\end{figure}

LaBi has a NaCl-type centrosymmetric structure with the symmorphic space group Fm-3m. The Wyckoff position of La and Bi are 4$a$ (0, 0, 0) and 4$b$ (1/2, 1/2, 1/2), respectively. The XRD pattern of a LaBi single crystal indicates that the surface of crystal is the (l 0 0) plane (figure 1(a)). The cubic-shape LaBi crystals (inset of figure 1(a)) is consistent with the single crystal XRD pattern and its crystallographic symmetry. As shown in figure 1(b), the zero-field resistivity $\rho_{xx}(T,0)$ of LaBi single crystal exhibits metallic behavior, similar to that in LaSb \cite{Tafti}. The residual resistance ratio [RRR $\equiv R$(300 K)/$R$(2 K) = 204] indicates the high quality of sample. The linear temperature dependence of $\rho_{xx}(T,0)$ at high temperatures suggests the electron-phonon ($e$-$ph$) scattering is the dominant scattering mechanism. In contrast, the $\rho_{xx}(T,0)$ at low-temperature region (2 K $\leq T \leq$ 20 K) can be well fitted by using the formula $\rho_{xx}(T,0)=\rho_{0}+AT^{n}$ with $\rho_{0}=$ 0.1765(3) $\mu\Omega$ cm, $A=$ 2.528(4)$\times$10$^{-5}$ $\mu\Omega$ cm K$^{-3}$ and $n=$ 2.99(1) (inset of figure 1(b)). The value of $\rho_{0}$ is larger than that in LaSb ($\rho_{0}=$ 0.08 $\mu\Omega$ cm) but the $n$ is smaller ($n=$ 4 for LaSb) \cite{Tafti}. The values of $n$ for both systems distinctly different from $n=$ 2 when $e$-$e$ scattering is dominant, such as in WTe$_{2}$ \cite{WangYL}. These values are also smaller than $n=$ 5 for the conventional $e$-$ph$ scattering process according to the Bloch-Gr\"{u}neisen theory \cite{Ziman}. It suggests that the interband $s$-$d$ $e$-$ph$ scattering, rather than the intraband $s$-$s$ $e$-$ph$ scattering should be dominant at low temperature \cite{Ziman}. Similar behavior has been observed in yttrium metal as well as in other transition metal carbide \cite{Arajs,ZhangX}.

Figure 1(c) shows the $\rho_{xx}(T, \mu_{0}H)$ at various fields up to 14 T. When the field is low, the $\rho_{xx}(T, \mu_{0}H)$ still exhibits metallic behavior in the whole temperature region, similar to the $\rho_{xx}(T, 0)$. Once the field is beyond the critical field $\mu_{0}H_{c}$ ($\sim$ 0.5 T) (inset of figure 1(c)), the $\rho_{xx}(T, \mu_{0}H)$ curves show a minimum at field-dependent "turn-on" temperatures $T_{m}(\mu_{0}H)$, i.e., the $\rho_{xx}(T, \mu_{0}H)$ increases significantly with decreasing temperature. This behavior has been observed not only in LaSb but also in other XMR materials, such as WTe$_{2}$ and NbP \cite{Shekhar,Mazhar,Tafti}. Correspondingly, when $\mu_{0}H>\mu_{0}H_{c}$, the $d\rho_{xx}(T, \mu_{0}H)/dT$ is positive at $T>T_{m}$ and changes the sign at $T<T_{m}$ (inset of figure 1(d)). When the field increases, the $T_{m}$ shifts to higher temperature, but another characteristic temperature $T_{i}$ related to the inflection point of $\rho_{xx}(T, \mu_{0}H)$ seems insensitive to the field (figure 1(d)). Both characteristic temperatures merge together when $\mu_{0}H\sim \mu_{0}H_{c}$. Moreover, there is a plateau in $\rho_{xx}(T, \mu_{0}H)$ curves following the upturn behavior at low temperatures and high fields. Similar behaviors have been observed in LaSb \cite{Tafti}.

\begin{figure}[tbp]
\centerline{\includegraphics[scale=0.6]{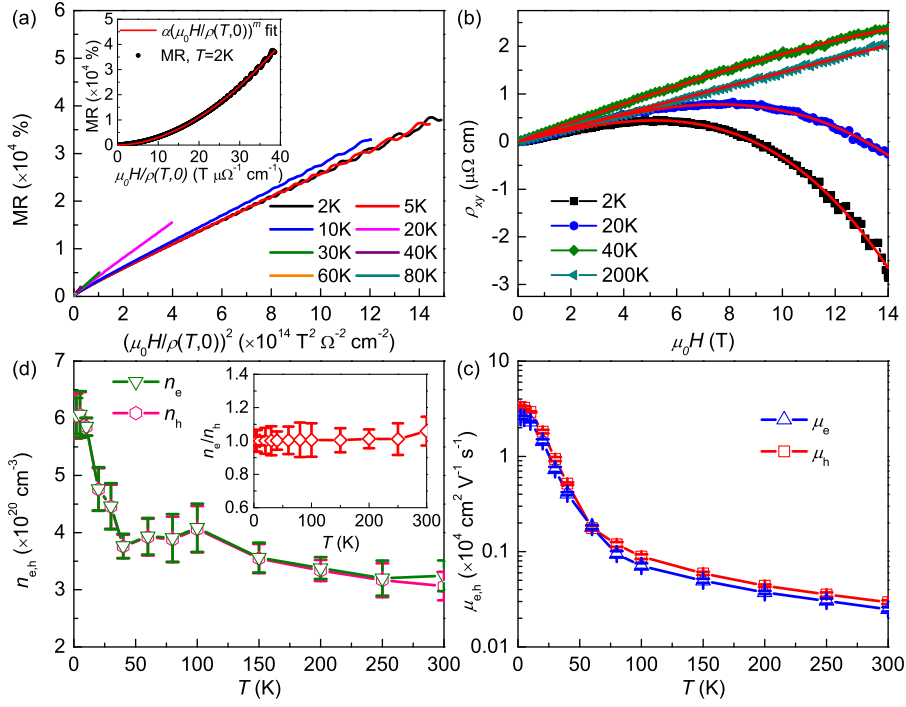}} \vspace*{-0.3cm}
\caption{(a) MR vs. $[\mu_{0}H/\rho_{xx}(T, 0)]^{2}$ at $T\leq$ 80 K. Inset: MR at 2 K. The solid line is the fit using $\alpha [\mu_{0}H/\rho_{xx}(T,0)]^{m}$. (b) Field dependence of $\rho_{xy}(T,\mu_{0}H)$ at various temperatures. The red solid lines are the fits using the two-band model. (c) Temperature dependence of fitted $n_{e,h}(T)$. Inset: the ratio of $n_{e}/n_{h}$ a function of $T$. (d) Temperature dependence of fitted $\mu_{e,h}(T)$.}
\end{figure}

LaBi exhibits rather large MR ($=(\rho_{xx}(T,\mu_{0}H)-\rho_{xx}(T,0))/\rho_{xx}(T,0)=\Delta\rho_{xx}(T,\mu_{0}H)/\rho_{xx}(T,0)$) at low temperature and the MR at 2 K is 3.8$\times$10$^{4}$ \% under 14 T (figure 2(a)). Although the MR at 9 T ($\sim$ 1.6$\times$10$^{4}$ \%) is one order of magnitude smaller than that in LaSb, it is still comparable with the MRs in other XMR materials \cite{Huang,LiangT,Tafti}. The MR of LaBi does not saturate up to 14 T and when the field is larger than about 5 T, the Shubnikov-de Haas (SdH) quantum oscillations (QOs) appears at low temperature (figure 2(a)). When $T\leq$ 10 K, the MR decreases gradually with increasing temperature, but drops quickly at higher temperatures. According to the Kohler's rule \cite{WangYL,Ziman}, MR $=\alpha(\mu_{0}H/\rho_{xx}(T,0))^{m}$, if the carrier concentration of sample is weakly temperature dependent, the MR measured at various temperatures can be scaled into a single curve. But the scaled MR curves do not fall into one curve for LaBi (figure 2(a)), obviously deviating from the Kohler's rule. This violation strongly suggests that the carrier concentrations and/or the mobility ratio of hole to electron change significantly with temperature \cite{WangYL,ZhanYF}. The fit of MR curve at 2 K gives $\alpha=$ 0.411(4) ($\mu\Omega$ cm T$^{-1}$)$^{1.866}$ and $m=$ 1.866(2) (inset of figure 2(a)), close to the typically quadratic field dependence of the MR in the multiband metals.

Hall resistivity $\rho_{xy}(T,\mu_{0}H)$ of LaBi (figure 2(b)) is positive at high temperature with nearly linear field dependence and when lowering temperature, the $\rho_{xy}(T,\mu_{0}H)$ exhibits nonlinear behavior and becomes negative at high field. Moreover, the QOs are clearly seen at 2 K, consistent with the MR results. The $\rho_{xy}(T,\mu_{0}H)$ can be fitted using the two-band model \cite{WangYL,Ziman},

\begin{equation}
\rho_{xy} = \frac{\mu_{0}H}{e}\frac{(n_{h}\mu_{h}^{2}-n_{e}\mu_{e}^{2})+(n_{h}-n_{e})(\mu_{e}\mu_{h})^{2}(\mu_{0}H)^{2}}{(n_{h}\mu_{h}+n_{e}\mu_{e})^{2}+(n_{h}-n_{e})^{2}(\mu _{e}\mu _{h})^{2}(\mu _{0}H)^{2}}
\end{equation}

where $n_{e,h}$ and $\mu_{e,h}$ are the carrier concentrations and mobilities of electron and hole, respectively. The fits are quiet well (solid lines in figure 2(b)) and the obtained $n_{e,h}(T)$ and $\mu_{e,h}(T)$ as a function of temperature are shown in figure 2(c) and (d). At $T>$ 40 K, the $n_{e,h}(T)$ is weakly temperature dependent. When lowering temperature, the $n_{e,h}(T)$ increase gradually at the beginning and then become saturated. The estimated carrier concentration at 2 K is 6.0(4) and 6.0(3)$\times$10$^{20}$ cm$^{-3}$ for $n_{e}$ and $n_{h}$, respectively. Such small $n_{e,h}$ with two types of carriers confirms the semimetallicity of LaBi. More importantly, the ratio of $n_{e}$/$n_{h}$ is almost one with slightly larger $n_{e}$ than $n_{h}$ in the whole temperature range, strongly suggesting that the carriers in LaBi is nearly compensated. On the other hand, the $\mu_{e,h}(T)$ monotonically increases with decreasing temperature with a larger slope between 10 - 40 K and becomes rather high at low temperature (2.6(1) and 3.3(1)$\times$10$^{4}$ cm$^{2}$ V$^{-1}$ s$^{-1}$ for $\mu_{e}$ and $\mu_{h}$ at 2 K, respectively). Strong temperature dependence of $n_{e,h}(T)$ and $\mu_{e,h}(T)$ between 10 - 40 K possibly causes the obvious violation of Kohler's rule in this temperature range (figure 2(a)).

\begin{figure}[tbp]
\centerline{\includegraphics[scale=0.25]{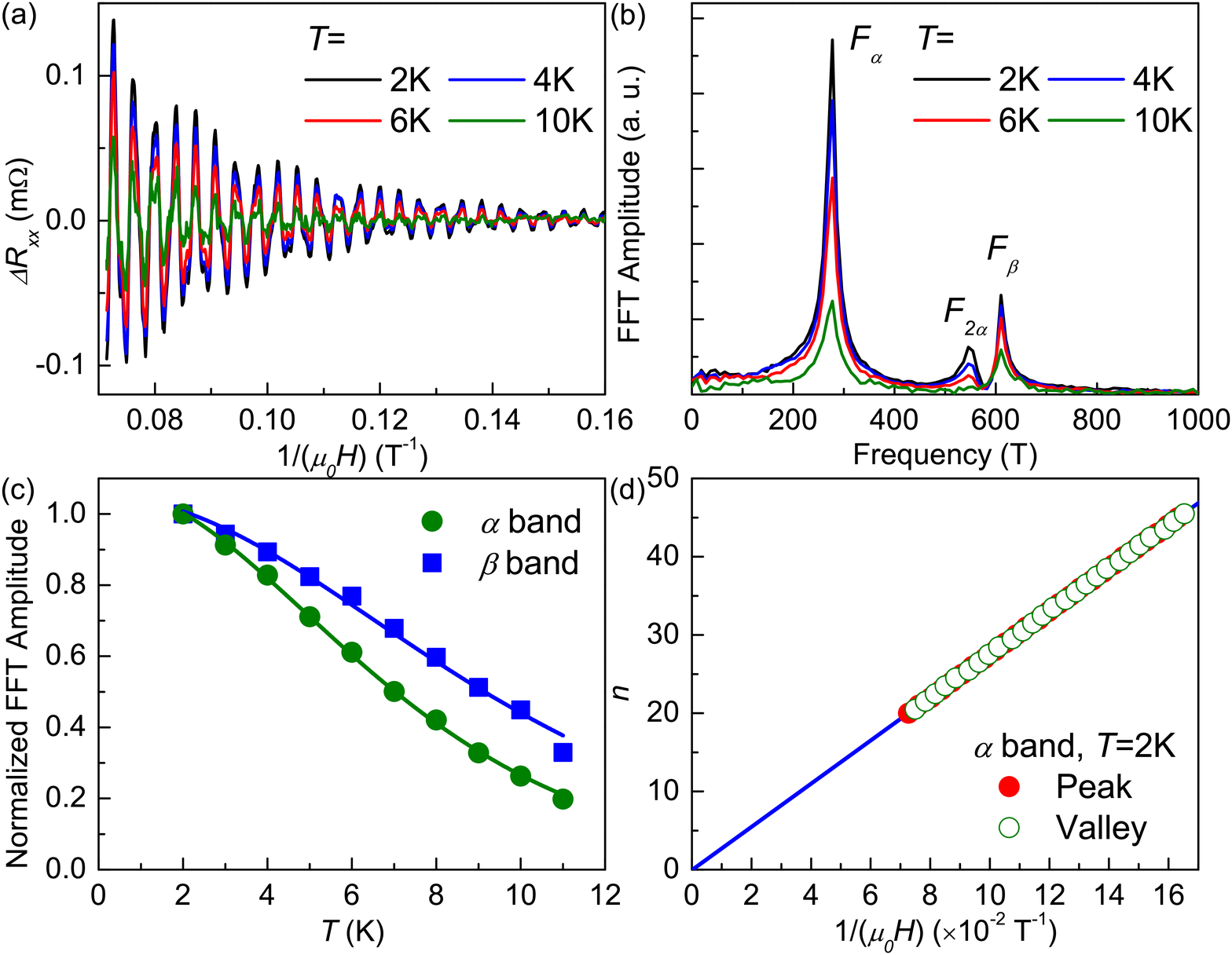}} \vspace*{-0.3cm}
\caption{(a) SdH QOs $\Delta R_{xx} = R_{xx}-<R_{xx}>$ as a function of 1/($\mu_{0}H$) at various temperatures. (b) FFT spectra of the QOs at various temperatures. (c) The temperature dependence of FFT amplitudes of $\alpha$ and $\beta$ peaks, normalized by their 2 K values. The solid lines are the fits using the LK formula, giving the effective masses. (d) The Landau level indices $n$ as a function of 1/($\mu_{0}H$) for the $\alpha$ band. The solid and empty symbols correspond to the positions of peaks and valleys in $\Delta R_{xx}$ curve at 2 K, respectively. The solid line represents the linear fit of data.}
\end{figure}

The QOs as a function of $1/(\mu_{0}H)$ at low temperatures are shown in figure 3(a). The amplitudes of QOs decrease with increasing temperature or decreasing field. The fast Fourier transform (FFT) spectra of the QOs reveals two principal frequencies $F_{\alpha}=$ 276 T with its second harmonic frequency $F_{2\alpha}=$ 555 T and $F_{\beta}=$ 610.5 T (figure 3(b)). All of these frequencies are higher than those in LaSb \cite{Tafti}, indicating LaBi has larger FSs than LaSb, agreeing with the results of Hall measurement \cite{Tafti}. According to the Onsager relation $F=(\hbar/2\pi e)A_{F}$, where $A_{F}$ is the extremal cross section of FS. The determined $A_{F}$ is 0.027 and 0.058 \AA$^{-2}$ for $\alpha$ and $\beta$ bands, respectively, consistent with previous results \cite{Hasegawa}. These $A_{F}$s are very small, taking only $\sim$ 2.9 and 6.4 \% of the whole area of Brillouin zone in the $k_{x}-k_{y}$ plane providing the lattice parameter $a=$ 6.58 \AA. Because the FS of $\beta$ band is almost spherical \cite{Hasegawa}, and there is only one $\beta$ FS in one Brillouin zone according to the band structure calculation shown below, the estimated carrier concentration for $\beta$ band $n_{h}^{\beta}$ is 0.85$\times$10$^{20}$ cm$^{-3}$. The effective mass $m^{*}$ can be extracted from the temperature dependence of the amplitude of FFT peak using the Lifshitz-Kosevich (LK) formula \cite{Shoenberg},

\begin{equation}
\Delta\rho_{xx} \propto \frac{X}{\sinh X}e^{-2\pi^{2}k_{B}T_{D}/\hbar \omega_{c}}\cos 2\pi(F/(\mu_{0}H)+1/2-\beta+\delta)
\end{equation}

where $X=2\pi^{2}k_{B}T/\hbar\omega_{c}=14.69m^{*}/\mu_{0}H_{avg}$ with $\hbar\omega_{c}$ being the cyclotron frequency and $\mu_{0}H_{avg}$ [= (14 - 5.44)/2 = 9.72 T] being the average value of the field window used for the FFT spectra of QOs \cite{Rhodes}. $T^{\alpha}_{D}$ is the Dingle temperature, $\beta$ is the Berry's phase, and $\delta$ is a phase shift determined by the dimensionality ($\delta=\pm$ 1/8 for the 3D system) \cite{Shoenberg,Mikitik,Lukyanchuk}. The fitted $m^{*}$ of $\alpha$ and $\beta$ band is 0.217(1) and 0.164(3) $m_{e}$, respectively, where $m_{e}$ is the bare electron mass (figure 3(c)). The fitted $T_{D}$ from the decaying amplitude of QOs with field at 2 K \cite{Shoenberg} gives 9(1) K for the $\alpha$ band, corresponding to the quantum lifetime $\tau^{\alpha}_{Q}=\frac{\hbar}{2\pi k_{B}T^{\alpha}_{D}}=$ 1.2(2)$\times$10$^{-13}$ s. According to eq. (2), the Landau Level (LL) index $n$ is related to the QO frequency $F$ by the Lifshitz-Onsager quantization rule $F/(\mu_{0}H)=n+1/2-\beta+\delta$ \cite{Shoenberg,Mikitik,Lukyanchuk}. The peaks and valleys of the $\Delta R_{xx}$ at 2 K are assigned as integer ($n$) and half integer ($n+1/2$) LL indices, respectively. Linear fit of $n$ vs. $1/(\mu_{0}H)$ yields $F=$ 275.9(4) T and $1/2-\beta+\delta=$ -0.05(5) (figure 3(d)). For normal metals with trivial parabolic dispersion relation, the Berry's phase $2\pi\beta=$ 0. For Dirac systems with linear dispersion, there should be a nontrivial $\pi$ Berry's phase ($\beta=$ 1/2), as observed in 2D graphene and 3D Dirac semimetal Cd$_{3}$As$_{2}$ \cite{Novoselov, HeLP}. The obtained $\beta=0.55(5)+2\pi\delta$ indicates a possible nontrivial $\pi$ Berry's phase for the $\alpha$ band of LaBi. It should be noted that because the lowest LL obtained at present is $n=$ 7 which is far from the quantum limit, there might be a relatively large uncertainty when determining the intercept. This nontrivial $\pi$ Berry's phase needs to be further confirmed by the measurement at much higher field in order to be closer the quantum limit.

\begin{figure}[tbp]
\centerline{\includegraphics[scale=0.5]{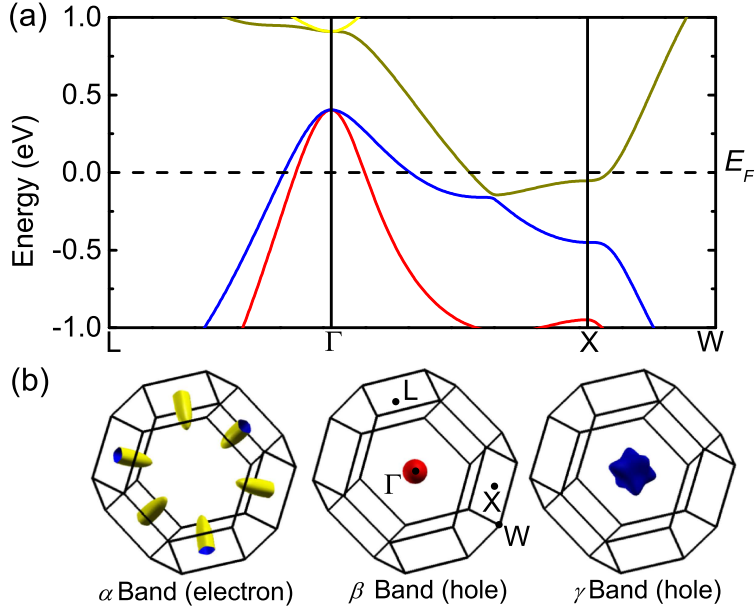}} \vspace*{-0.3cm}
\caption{(a) Band structure and (b) FSs of LaBi calculated with the SOC effect. The dashed line denotes the position of $E_{F}$. There are three FSs at $E_{F}$, including two hole-type FSs around $\Gamma$ point and one electron-type FS around $X$ point.}
\end{figure}

Similar to LaSb \cite{Tafti,Zeng}, theoretical calculation indicates that there are three bands crossing $E_{F}$ in LaBi (figure 4(a)), consistent with QOs results and previous calculations \cite{Hasegawa}. The electron-type band ($\alpha$ band) is located at $X$ point with the ellipsoidal FS with longest principle axis along $\Gamma-X$ line; Two hole-type bands ($\beta$ and $\gamma$ bands) are centered at $\Gamma$ point. One has a nearly spherical FS and another one has a FS stretched in the $<100>$ directions (figure 4(b)). The calculated $n_{e,h}$ are 3.52, 0.72, and 2.88$\times$10$^{20}$ cm$^{-3}$ for the $\alpha$, $\beta$, and $\gamma$ bands, strongly indicating that LaBi is a nearly compensated SM ($\lambda=n_{e}$/$n_{h}=$ 0.98).

\begin{figure}[tbp]
\centerline{\includegraphics[scale=0.6]{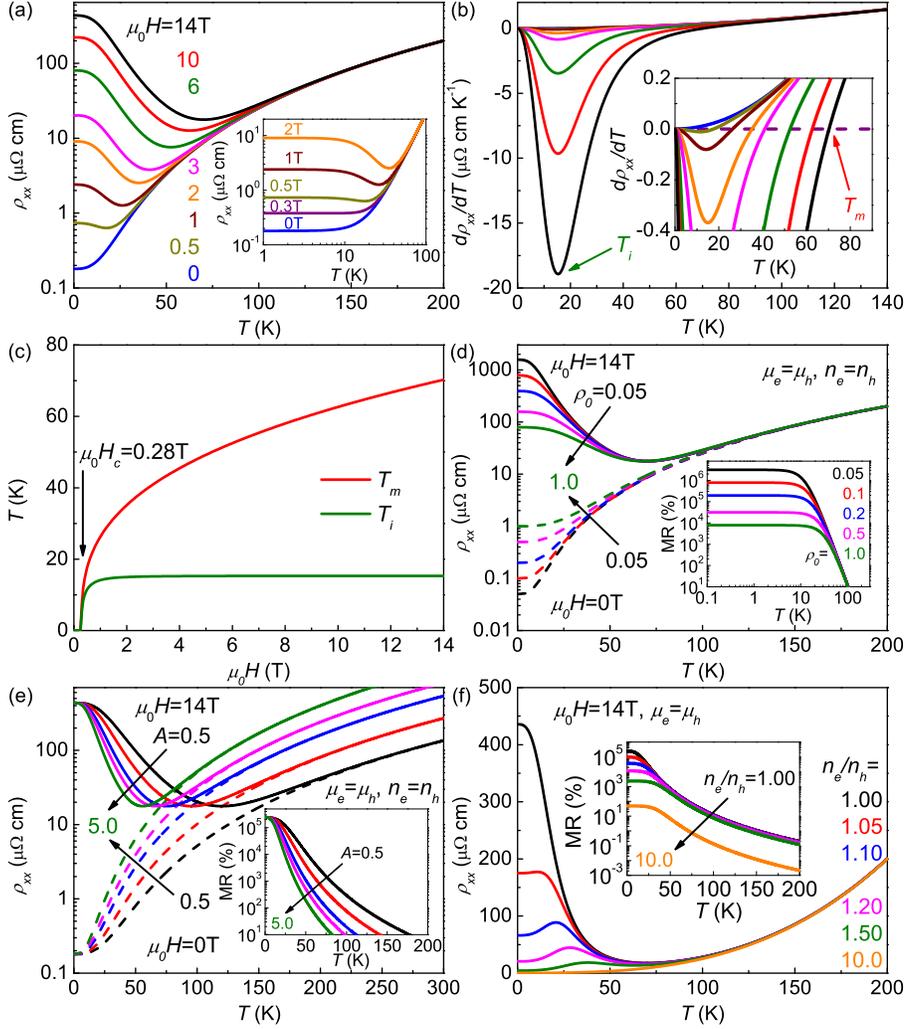}} \vspace*{-0.3cm}
\caption{Simulation of longitudinal resistivity with $\rho_{0}=$ 0.18 $\mu\Omega$ cm, $A=$ 2.5$\times$10$^{-5}$ $\mu\Omega$ cm K$^{-3}$, $n=$ 3, and $\alpha=$ 0.4 ($\mu\Omega$ cm T$^{-1}$)$^{2}$. (a) $\rho_{xx}(T, \mu_{0}H)$ vs. $T$ at various fields with $\mu_{0}H_{c}=$ 0.28 T. Inset: simulated $\rho_{xx}(T, \mu_{0}H)$ vs. $T$ when the fields are near $\mu_{0}H_{c}$. (b) Corresponding $d\rho_{xx}(T, \mu_{0}H)/dT$ as a function of $T$ at various fields. Inset displays the changes of $T_{m}$ with field. The color code is same as in (a). (c) Field dependence of derived $T_{m}$ and $T_{i}$. (d) $\rho_{xx}(T, \mu_{0}H)$ vs. $T$ at $\mu_{0}H=$ 0 (dashed lines) and 14 T (solid lines) with $\rho_{0}=$ 0.05, 0.1, 0.2, 0.5, and 1.0 $\mu\Omega$ cm. (e) $\rho_{xx}(T, \mu_{0}H)$ vs. $T$ at $\mu_{0}H=$ 0 (dashed lines) and 14 T (solid lines) with $A=$ 0.5, 1, 2, 3, and 5$\times$10$^{-5}$ $\mu\Omega$ cm K$^{-3}$. (f) $\rho_{xx}(T, \mu_{0}H)$ vs. $T$ at various $n_{e}/n_{h}$ with $\mu_{0}H=$ 14 T. Insets of (d)-(f): temperature dependence of corresponding MR at $\mu_{0}H=$ 14 T.}
\end{figure}

Because LaSb and LaBi seem nearly compensated SMs, similar to WTe$_{2}$, we develop the model proposed to explain the mangetotransport properties of WTe$_{2}$ \cite{WangYL} to analysis the common features in (nearly) compensated SMs. According to the two-band model \cite{WangYL,Ziman}, the MR can be expressed

\begin{equation}
\mathrm{MR} =\frac{n_{e}\mu_{e}n_{h}\mu_{h}(\mu_{e}+\mu_{h})^{2}(\mu_{0}H)^{2}}{(n_{e}\mu_{e}+n_{h}\mu_{h})^{2}+(n_{e}-n_{h})^{2}(\mu _{e}\mu _{h})^{2}(\mu _{0}H)^{2}}
\end{equation}

for the compensated SMs ($n_{e}=n_{h}=n_{c}$), we have MR $=\mu_{e}\mu_{h}(\mu_{0}H)^{2}$. Thus, the large MR is directly related to the high mobilities of materials. If assuming $\rho_{xx}(T,0)=\rho_{0}+AT^{n}$, we get

\begin{equation}
\rho_{xx}(T,\mu_{0}H)=\rho_{0}+AT^{n}+\alpha(\mu_{0}H)^{2}/(\rho_{0}+AT^{n})
\end{equation}

where $\alpha=\kappa[n_{c}e(1+\kappa)]^{-2}$ and $\kappa=\mu_{h}/\mu_{e}$ \cite{WangYL}. Therefore, we will return to the Kohler's rule with $m=$ 2. More importantly, according to eq. (4), there is a minimum in $\rho_{xx}(T,\mu_{0}H)$ curve when $\mu_{0}H>\mu_{0}H_{c} (=\rho_{0}/\alpha^{1/2})$ \cite{WangYL}, i.e., field-induced resistivity upturn. The simulated $\rho_{xx}(T,\mu_{0}H)$ curves with $\mu_{0}H_{c}=$ 0.28 T clearly reflect this behavior and fairly well reproduce the experimental results (figure 5(a)). Firstly, the $\rho_{xx}(0,\mu_{0}H)$ is not infinity but $\rho_{0}+\alpha(\mu_{0}H)^{2}/\rho_{0}$, i.e., there is a resistivity plateau at low temperature with the absolute value increasing at higher field. Secondly, the simulated field dependence of $T_{m}$ and $T_{i}$ well reflect the features observed in LaSb \cite{Tafti} and LaBi (figure 5(b) and (c)). The $T_{m}$ is proportional to $(\mu_{0}H-\mu_{0}H_{c})^{1/n}$ (here, $n=$ 3 for LaBi) and the $T_{i}$ saturates quickly when field is large. Noted that both $T_{i}(\mu_{0}H_{c})$ and $T_{m}(\mu_{0}H_{c})$ become zero but these trend are not observed experimentally, possibly because the initial slope of $T_{i}(\mu_{0}H_{c})$ and $T_{m}(\mu_{0}H_{c})$ are very large and the $\rho_{xx}(T,\mu_{0}H)$ is too flat to extract accurate $T_{i}$ and $T_{m}$ when $\mu_{0}H$ is close to $\mu_{0}H_{c}$.

Thirdly, if assuming $\mu_{e}=\mu_{h}=\mu$ and temperature-independent $n_{e}=n_{h}=n_{c}$, it has $\rho=\sigma^{-1}=(n_{e}e\mu_{e}+n_{h}e\mu{h})^{-1}=(2n_{c}e\mu)^{-1}=\rho_{0}+AT^{n}$, and thus when $\mu$ increases, it corresponds to the decrease of either $\rho_{0}$ and/or $A$. Figure 5(d) shows the influence of $\rho_{0}$ on magnetotransport behavior. It can be seen that with smaller $\rho_{0}$, the upturn behavior becomes more obvious. Moreover because the increase of $\rho_{xx}(T,\mu_{0}H)$ at high field is accompanied by the decrease of $\rho_{xx}(T,0)$ , the MR at low temperature region is remarkably influenced by the $\rho_{0}$ (inset of figure 5(d)). The small $\rho_{0}$, partly originating from the large $\mu$, will lead to much obvious MR. It is consistent with the results in terms of eq. (3) under compensation condition. On the other hand, if $\rho_{0}$ is fixed, i.e., $\mu(0)=$ constant, the smaller $A$ leads to the weaker decay of $\mu(T)$ with increasing temperature, and thus the upturn behavior appears higher temperature region and becomes more obvious (figure 5(e)). Correspondingly, the MR at same temperature and field becomes large with smaller $A$ (inset of figure 5(e)). Overall, the increase of $\mu$ is in favor of field-induced upturn behavior and large MR.

Finally, we examine the influence of the mobilities on the uncompensation between electron and hole on the field-induced upturn behavior. If assuming $\mu_{e}=\mu_{h}$, we get

\begin{equation}
\rho_{xx}(T,\mu_{0}H)=\rho_{0}+AT^{n}+\frac{4\lambda(\rho_{0}+AT^{n})(\mu_{0}H)^{2}}{\beta^{2}(\rho_{0}+AT^{n})^{2}(\lambda+1)^{4}+(\lambda-1)^{2}(\mu_{0}H)^{2}}
\end{equation}

where $\beta=\frac{1}{2\sqrt{\alpha}}$. When $\lambda(=n_{e}$/$n_{h})$ increases, although the $\rho_{xx}(0,\mu_{0}H)$ decreases gradually, there is still a remarkable resistivity upturn with a plateau even extending to the higher temperature (figure 5(d)). It indicates that the field-induced resistivity upturn is robust to the imbalance between $n_{e}$ and $n_{h}$. Surprisingly, a reentrant metallic behavior is observed with further increasing $\lambda$. It is because the second item of the denominator in eq. (5), originating from the uncompensated carriers, weakens the effect of field on $\rho_{xx}(T,\mu_{0}H)$ at low temperature. Correspondingly, the MR also decreases with increasing $\lambda$ (inest of figure 5(f)). In the extreme case that the $n_{e}$ and $n_{h}$ are severely imbalanced, the upturn behavior even disappears (Fig. 5(d)) since the system will become a (nearly) single-band system where the MR is very small, i.e., the field does not have effect on $\rho_{xx}(T,\mu_{0}H)$. We have to mention that above analysis is valid when $n\geq$ 2.

\begin{figure}[tbp]
\centerline{\includegraphics[scale=0.5]{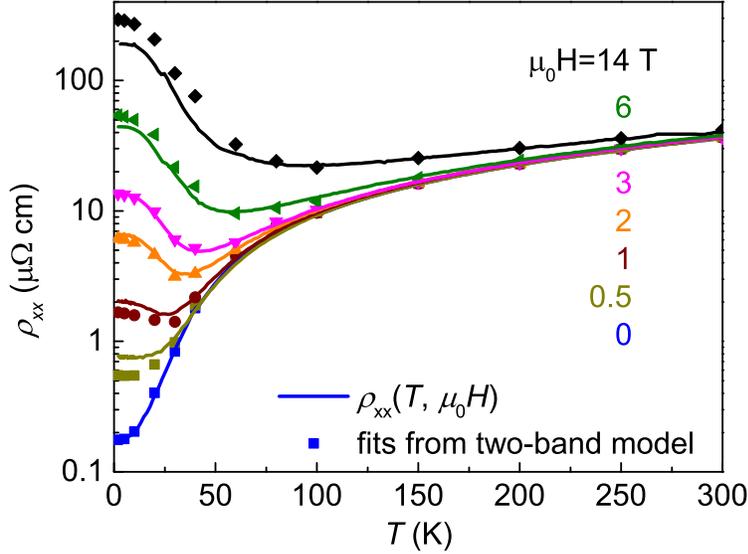}} \vspace*{-0.3cm}
\caption{Temperature dependence of $\rho_{xx}(T, \mu_{0}H)$ at various fields (solid lines). The scatters are calculated using the temperature dependent $n_{e,h}$ and $\mu_{e,h}$ derived from the two-band model analysis.}
\end{figure}

In order to further confirm the validity of above analysis, the calculated $\rho_{xx}(0,\mu_{0}H)$ using the temperature dependent $n_{e,h}$ and $\mu_{e,h}$ derived from the two-band model analysis is plotted together with the measured $\rho_{xx}(0,\mu_{0}H)$ (figure 6). It can be seen that both of results are consistent with each other very well and similar analysis was carried out in classic semimetallic bismuth and graphite \cite{DuX}. It strongly implies that the origin of field-induced properties in LaBi, similar to other semimetals with compensated carriers, can be well understood in the framework of multiband model. Noted that recently discovered TaSb$_{2}$ \cite{LiYK} seems to exhibit similar behaviors as in La$Pn$. Thus, it would be interest to examine if our analysis will apply to this system.

Except the electron-hole compensation mechanism, the magnetic field lifting of topologically protected backscattering in topological SMs could also lead to XMR \cite{LiangT}. In these systems, the backscattering is strongly suppressed at zero field, resulting in a transport lifetime $\tau_{tr}$ much longer than the quantum lifetime $\tau_{Q}$. We check this possible mechanism of XMR in LaBi. The $\tau_{tr}=m^{*}\mu_{e,h}/e$ is 2.59(6)$\times$10$^{-12}$ s and the ratio $\tau_{tr}/\tau^{\alpha}_{Q}$ ($\sim$ 22) is much smaller than those in Dirac SM Cd$_{3}$As$_{2}$ ($\sim$ 10$^{4}$) and Weyl SM NbAs ($\sim$ 10$^{3}$) \cite{LuoY,LiangT}, implying that there may be little topological protection for the $\alpha$ band of LaBi.

\section{Conclusion}

In summary, LaBi exhibits the XMR and field-induced resistivity upturn, similar to isostructural LaSb. Based on the two-band model, these behaviors can be well understood and the electron-hole compensation mechanism seems dominant in LaBi. The key conditions leading to the XMR and field-induced properties in SMs are \emph{significantly high $\mu_{e,h}$, extremely small $\rho_{0}$, and rather low $n_{e,h}$ satisfying the compensation condition}. Because theoretical calculation suggests that LaP and LaAs have smaller overlap between electron- and hole-type bands with lower $n_{e,h}$ than LaSb \cite{Zeng}, it is worth inspecting if they will exhibit even larger MR. Systematical studies on La$Pn$ family with simple electron structure will not only deepen our understanding on XMR phenomenon but also shed light on exploring novel XMR materials with prominent properties for potential applications.

\ack{
This work was supported by the Ministry of Science and Technology of China (973 Project: 2012CB921701), the Fundamental Research Funds for the Central Universities, and the Research Funds of Renmin University of China (14XNLQ03 and 15XNLF06), and the National Nature Science Foundation of China (Grant No. 11574394). Computational resources have been provided by the Physical Laboratory of High Performance Computing at RUC.}

$\dag$ These authors contributed equally to this work.

\end{document}